\documentclass[journal]{IEEEtran}

\usepackage{color,array,amsthm}
\usepackage{graphicx}
\usepackage{amsmath,amsfonts}
\usepackage{siunitx}
\usepackage{booktabs}
\usepackage{multirow}
\usepackage{balance}
\usepackage{url}
\usepackage{cite}
\usepackage[caption=false,font=footnotesize]{subfig}

\begin{document}

\title{Inter-Year Transfer of Altitude-Dependent Spectrum Activity Models Using Minimal Calibration}

\author{Amir~Hossein~Fahim~Raouf,~and~\.{I}smail~G\"uvenc%
\thanks{A. H. F. Raouf and \.{I}. G\"uvenc are with the Department of Electrical and Computer Engineering, North Carolina State University, Raleigh, NC, USA (e-mail: amirh.fraouf@ieee.org; iguvenc@ncsu.edu).}%
\thanks{This work was supported in part by the National Science Foundation under Grant CNS-2332835.}%
}

\maketitle

\begin{abstract}
This paper studies the transferability of altitude-dependent spectrum activity models and measurements across years. We introduce a physics-informed, mean-only stochastic-geometry model of aggregate interference to altitude-binned received power, yielding three interpretable parameters for a given band and campaign: 1) line-of-sight transition slope, 2) transition altitude, and 3) effective activity constant. 
Analysis of aerial spectrum measurements collected from 2023 to 2025 across multiple sub-6~GHz bands reveals that downlink~(DL) and shared-access bands preserve a persistent geometry-driven altitude structure that is stable across years. In contrast, uplink~(UL) bands exhibit weak altitude dependence with no identifiable transition, indicating that interference is dominated by activity dynamics rather than propagation geometry. To quantify the practical limits of model reuse, we evaluate a minimal-calibration method in which the transition altitude is fixed from a reference year and the remaining parameters are estimated from only two altitude bins in the target year. The results further indicate that the proposed approach provides accurate predictions for DL and CBRS bands, suggesting the feasibility of low-cost model transfer in stable environments, while highlighting the reduced applicability of mean-field models for UL scenarios.
\end{abstract}

\begin{IEEEkeywords}
UAV, aerial spectrum measurements, stochastic geometry, interference modeling, altitude dependence, transfer calibration.
\end{IEEEkeywords}

\section{Introduction}
Accurate characterization of altitude-dependent radio environments is a prerequisite for aerial coexistence analysis, interference assessment, and the design of multilayer wireless systems involving terrestrial and aerial nodes~\cite{Mozaffari2019Tutorial, Fotouhi2019Survey}. Although three-dimensional~(3D) spectrum maps can be constructed from large-scale measurement campaigns, repeating such campaigns across multiple years, frequency bands, and evolving network deployments is costly and often infeasible~\cite{Fotouhi2019Survey}. Aerial spectrum measurement platforms offer a scalable and flexible alternative. However, converting these measurements into compact, physically interpretable, and transferable models remains a fundamental challenge.

Empirical aerial measurements consistently reveal a non-monotonic altitude dependence in received power~\cite{AlHourani2014Optimal, maeng2025altitude, Azari2019Cellular}. At low altitudes, signal strength typically increases as line-of-sight~(LoS) probability improves and shadowing diminishes. Beyond a critical altitude, however, the mean received power decays due to expanding coverage footprints, increased path lengths, and aggregation over a larger set of signal sources. Mean-only stochastic geometry models provide a principled framework for capturing this behavior using a small number of physically meaningful parameters, such as effective path-loss exponents and altitude-dependent activity terms, without requiring detailed knowledge of instantaneous network states~\cite{Haenggi2012SG, Azari2019Cellular}.

A key open question is whether these parameters, once estimated from a given measurement campaign, remain transferable across years under realistic propagation variability and network evolution. Existing studies have largely emphasized in-sample fitting accuracy within individual datasets, implicitly assuming temporal stability of the inferred parameters~\cite{matolak2017air, Azari2019Cellular, maeng2025altitude}. In practice, long-term changes in infrastructure, traffic patterns, and environmental conditions introduce non-stationarities that may fundamentally limit parameter reuse.

This correspondence investigates the inter-year transferability of a physics-informed mean-field stochastic geometry model for aggregate interference derived from aerial spectrum measurements. Using data calibrated in one measurement campaign, we evaluate whether the model can be applied to subsequent years with minimal recalibration. Specifically, we investigate the identifiability and temporal stability of altitude-dependent propagation parameters and assess whether a two-point altitude calibration is sufficient to maintain predictive accuracy.

\section{Mean-Only Altitude-Dependent Interference Model}
\looseness=-1
Let $\Phi \subset \mathbb{R}^2$ denote a homogeneous Poisson point process~(PPP) of active ground transmitters with density $\lambda$ \cite{Haenggi2012SG}. Although terrestrial deployments are not strictly Poisson, a homogeneous PPP provides an effective mean-field approximation for spatially and temporally averaged aggregate interference and is sufficient for first-moment analysis~\cite{Haenggi2012SG}. We consider an unmanned aerial vehicle~(UAV) receiver located at altitude $H>0$ above the origin. A guard radius $r_0$ excludes transmitters whose horizontal distance from the UAV projection is less than $r_0$.

For a transmitter located at horizontal distance $r$, the three-dimensional link distance is
$R=\sqrt{r^2+H^2}$.
Rather than modeling LoS and non-line-of-sight~(NLoS) links as separate point processes, we adopt a \emph{mean-only equivalent representation} in which the effects of LoS probability, path-loss exponent, and shadowing variance are absorbed into altitude-dependent effective parameters. This modeling choice is motivated by prior air-to-ground measurement studies, which consistently report a smooth, non-monotonic dependence of mean received power on altitude rather than abrupt regime transitions~\cite{AlHourani2014Optimal, maeng2025altitude}. Then, the received power contribution from a transmitter is modeled as
\begin{equation}
P_{\mathrm{rx}}(r,H)=C_{\mathrm{eff}}\,S(H)\,R^{-\alpha(H)},
\end{equation}
where $C_{\mathrm{eff}}$ is an effective activity constant, $\alpha(H)$ is an altitude-dependent path-loss exponent, and $S(H)$ is a lognormal shadowing factor.

The altitude dependence of $\alpha(H)$ is governed by a smooth LoS transition model. Following commonly adopted air-to-ground channel models~\cite{AlHourani2014Optimal}, the probability of LoS is expressed as a logistic function of altitude,
\begin{equation}
p_{\mathrm{L}}(H)=\frac{1}{1+\exp[-\beta(H-H_0)]},
\end{equation}
where $\beta$ controls the transition sharpness and $H_0$ denotes the altitude at which LoS and NLoS conditions are equiprobable.
Using this probability, the path-loss exponent is interpolated between fixed LoS and NLoS values,
\begin{equation}
\alpha(H)=\alpha_{\mathrm{N}}-(\alpha_{\mathrm{N}}-\alpha_{\mathrm{L}})\,p_{\mathrm{L}}(H),
\end{equation}
where $\alpha_{\mathrm{L}}$ and $\alpha_{\mathrm{N}}$ denote the path-loss exponents corresponding to LoS and NLoS propagation conditions, respectively.
Shadowing is modeled as a unit-mean lognormal factor $S(H)$ such that
$\mathbb{E}[S(H)] = 1$. Shadowing variance affects only higher-order statistics and does not enter the mean expression. Consequently, any constant power offset is absorbed into $C_{\mathrm{eff}}$.
The aggregate received power at altitude $H$ is then given by
\begin{equation}\label{eq:agg_rec_pow}
I(H)=\sum_{x\in\Phi,\,\|x\|\ge r_0}
C_{\mathrm{eff}}\,S(H)\,R(x)^{-\alpha(H)}.
\end{equation}
\noindent Taking expectation of \eqref{eq:agg_rec_pow} with respect to the point process~$\Phi$ and applying Campbell's theorem~\cite{Haenggi2012SG}, we have
\begin{equation}\label{eq:mu_h}
\mu(H)=2\pi\lambda C_{\mathrm{eff}} 
\int_{r_0}^{\infty}(r^2+H^2)^{-\alpha(H)/2}\,r\,\mathrm{d}r.
\end{equation}
\noindent For $\alpha(H)>2$, the integral in~\eqref{eq:mu_h} converges to:
\begin{equation}\label{eq:mu_h_closed}
\mu(H)=
2\pi\lambda C_{\mathrm{eff}} 
\frac{(H^2+r_0^2)^{1-\alpha(H)/2}}{\alpha(H)-2}.
\end{equation}

The closed-form expression in~\eqref{eq:mu_h_closed} defines the mean-only stochastic geometry model, which we use throughout this work. The model captures altitude dependence exclusively through a smooth altitude-dependent LoS transition, which modulates the effective path-loss exponent in a physically interpretable manner. In contrast to approaches that rely on explicit LoS/NLoS point process thinning or higher-order interference statistics, the proposed formulation embeds LoS effects directly into the first-moment analysis, yielding a compact and analytically tractable representation consistent with prior mean-field UAV interference models~\cite{Chu2019Interference,Selim2025UAV}.

To enable cross-band comparison, we introduce a \textit{frequency-normalized} activity index, defined as
\begin{equation}
    \tilde{C} = C_{\mathrm{eff}} \left( \frac{4\pi f_c}{c} \right)^2 \frac{1}{G_t G_r},
\end{equation}
where $f_c$ is the center frequency of the band, $c$ is the speed of light, and $G_t$ and $G_r$ denote the transmitter and receiver antenna gains, respectively. This metric compensates for free-space path loss and antenna gain, yielding a measure of effective radiated activity that is independent of frequency. This ensures consistent comparisons across different bands and measurement years. For each campaign, we estimate the parameters $(\beta, H_0, \tilde{C})$ using a least-squares fit of the closed-form mean model to the altitude-binned data.

\section{Inter-Year Transfer Calibration}

Analysis of the spectrum measurements across multiple years reveals that while some fitted parameters remain approximately stable across measurement campaigns, others exhibit measurable drifts (see Table~\ref{tab:fit_summary}). To assess the impact of such drifts on model reuse, we consider a two-point transfer calibration in which only two altitude bins from a target year are used to calibrate the transition slope $\beta$ and effective activity factor $C_{\mathrm{eff}}$, while the break altitude $H_0$ is fixed to its value obtained from a reference campaign. 
Since the noise contribution is approximately constant with altitude, altitude-dependent trends are dominated by the interference component, and the model is interpreted as a mean interference model.

To calibrate the model using minimal data, we select two altitude bins, $H_1$ and $H_2$, with measured mean powers $Y_1$ and $Y_2$ from the target year. The predicted mean power $\mu(H)$ is factored into a magnitude term $C_{\mathrm{eff}}$ and a shape function $D$:
\begin{equation}
    \mu(H) = C_{\mathrm{eff}} \, D(H; \beta, H_0),
\end{equation}
where $D(H; \beta, H_0)$ encapsulates the altitude-dependent geometry and propagation effects.

To isolate the transition slope $\beta$, we form the log-ratio of the measurements, which eliminates the linear scaling factor $C_{\mathrm{eff}}$:
\begin{equation}
    \ln(Y_1)\! - \! \ln(Y_2) \! = \!  \ln\left( D(H_1; \beta, H_0) \right)\! - \! \ln\left( D(H_2; \beta, H_0) \right).
\end{equation}
We estimate $\beta$ by minimizing the squared error between the measured and predicted log-differences:
\begin{equation}
    \hat{\beta} = \arg\min_{\beta} \left| \ln\left(\frac{Y_1}{Y_2}\right) - \ln\left(\frac{D(H_1; \beta, H_0)}{D(H_2; \beta, H_0)}\right) \right|^2.
\end{equation}
Once $\hat{\beta}$ is obtained, $C_{\mathrm{eff}}$ is recovered by direct substitution:
\begin{equation}
    C_{\mathrm{eff}} = \frac{Y_1}{D(H_1; \hat{\beta}, H_0)}.
\end{equation}

This two-point procedure uses exactly two observations to identify two unknown parameters and therefore represents the minimal calibration required for inter-year transfer. It is not intended as a practical estimation strategy, but rather as a controlled stress test designed to expose identifiability limits and sensitivity to environmental drift.
The calibration altitudes are selected to bracket the transition region over which the altitude-dependent mean interference varies smoothly and reliably. In this paper, one low-to-mid altitude bin and one higher altitude bin are used to balance sensitivity to the transition slope against robustness to measurement noise and model mismatch.

\section{Numerical Results and Discussion}
For all numerical results, the mean-only stochastic geometry model is evaluated using fixed LoS and NLoS path-loss exponents $(\alpha_{\mathrm{L}},\alpha_{\mathrm{N}})=(2.2,3.6)$, transmitter density $\lambda=10^{-3}\,\mathrm{m}^{-2}$, guard radius $r_0=20\,\mathrm{m}$, unit antenna gains, and altitude bins spanning $H\in[10,160]\,\mathrm{m}$ with $5\,\mathrm{m}$ resolution.

\begin{table}[t]
\centering
\caption{Fitted altitude-dependent model parameters and goodness-of-fit metrics.}
\label{tab:fit_summary}
\scalebox{0.76}{
\begin{tabular}{l l c c c c c c}
\toprule
Year & Band & $\beta$ & $H_0$ [m] & $\tilde{C}$ & RMSE [dB] & $R^2_{\mathrm{lin}}$ & $R^2_{\mathrm{dB}}$ \\
\midrule
2023 & LTE B13 DL & 0.29 & 22.1 & 1.90 & 2.74 & 0.30 & 0.81 \\
2023 & LTE B13 UL & 0.01 & -199.6 & $6.21{\times}10^{-4}$ & 0.56 & 0.91 & 0.03 \\
2023 & 5G n5 DL & 0.08 & 14.0 & 1.23 & 2.77 & 0.87 & 0.85 \\
2023 & 5G n5 UL & 0.06 & -31.0 & $2.46{\times}10^{-3}$ & 0.68 & 0.35 & 0.41 \\
2023 & CBRS & 0.03 & -49.0 & $1.33{\times}10^{-3}$ & 0.35 & 1.00 & 0.90 \\
\midrule
2024 & LTE B13 DL & 0.35 & 27.0 & 1.00 & 2.97 & 0.81 & 0.78 \\
2024 & LTE B13 UL & 0.02 & -120.8 & $2.61{\times}10^{-3}$ & 0.25 & 0.68 & 0.22 \\
2024 & 5G n5 DL & 0.08 & 14.3 & 1.11 & 2.26 & 0.94 & 0.88 \\
2024 & 5G n5 UL & 0.05 & -35.5 & $5.47{\times}10^{-3}$ & 0.40 & 0.64 & 0.71 \\
2024 & CBRS & 0.05 & -14.3 & $1.38{\times}10^{-3}$ & 0.46 & 1.00 & 0.94 \\
\midrule
2025 & LTE B13 DL & 0.16 & 19.3 & 1.25 & 1.98 & 0.67 & 0.89 \\
2025 & LTE B13 UL & 0.03 & -101.5 & $2.35{\times}10^{-3}$ & 0.31 & 0.60 & 0.20 \\
2025 & 5G n5 DL & 0.08 & 23.7 & 2.48 & 1.95 & 0.94 & 0.94 \\
2025 & 5G n5 UL & 0.05 & -32.5 & $1.20{\times}10^{-2}$ & 0.71 & 0.40 & 0.48 \\
2025 & CBRS & 0.05 & -24.3 & $1.81{\times}10^{-3}$ & 0.44 & 1.00 & 0.90 \\
\bottomrule
\end{tabular}
}
\end{table}

\begin{figure*}[!t]
\centering
\newcommand{\figW}{0.29\linewidth}
\newcommand{\figH}{0.15\textheight}

\subfloat[5G n5 DL]{%
\includegraphics[width=\figW,height=\figH,keepaspectratio]{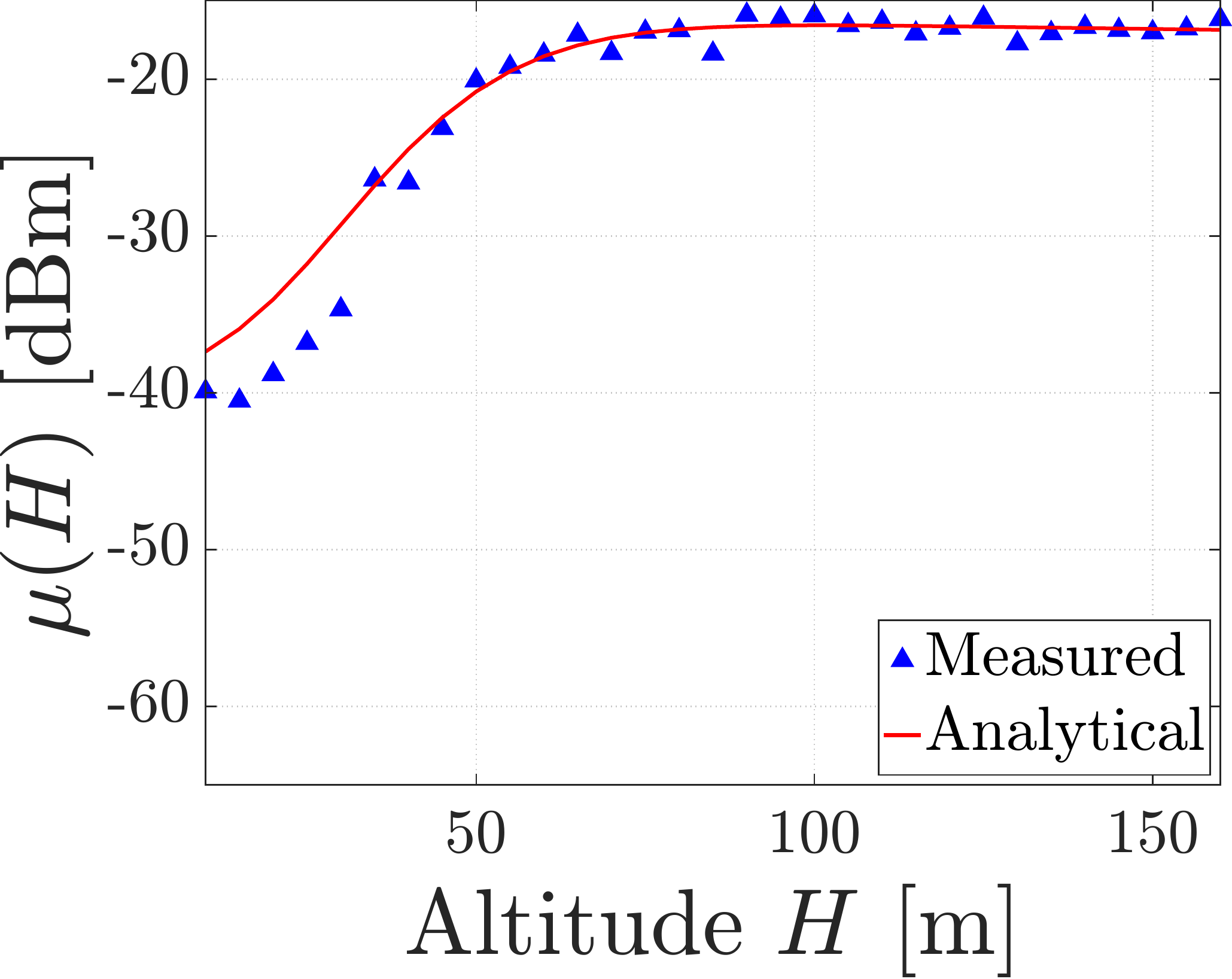}}
\hspace{0.01\linewidth}
\subfloat[5G n5 DL]{%
\includegraphics[width=\figW,height=\figH,keepaspectratio]{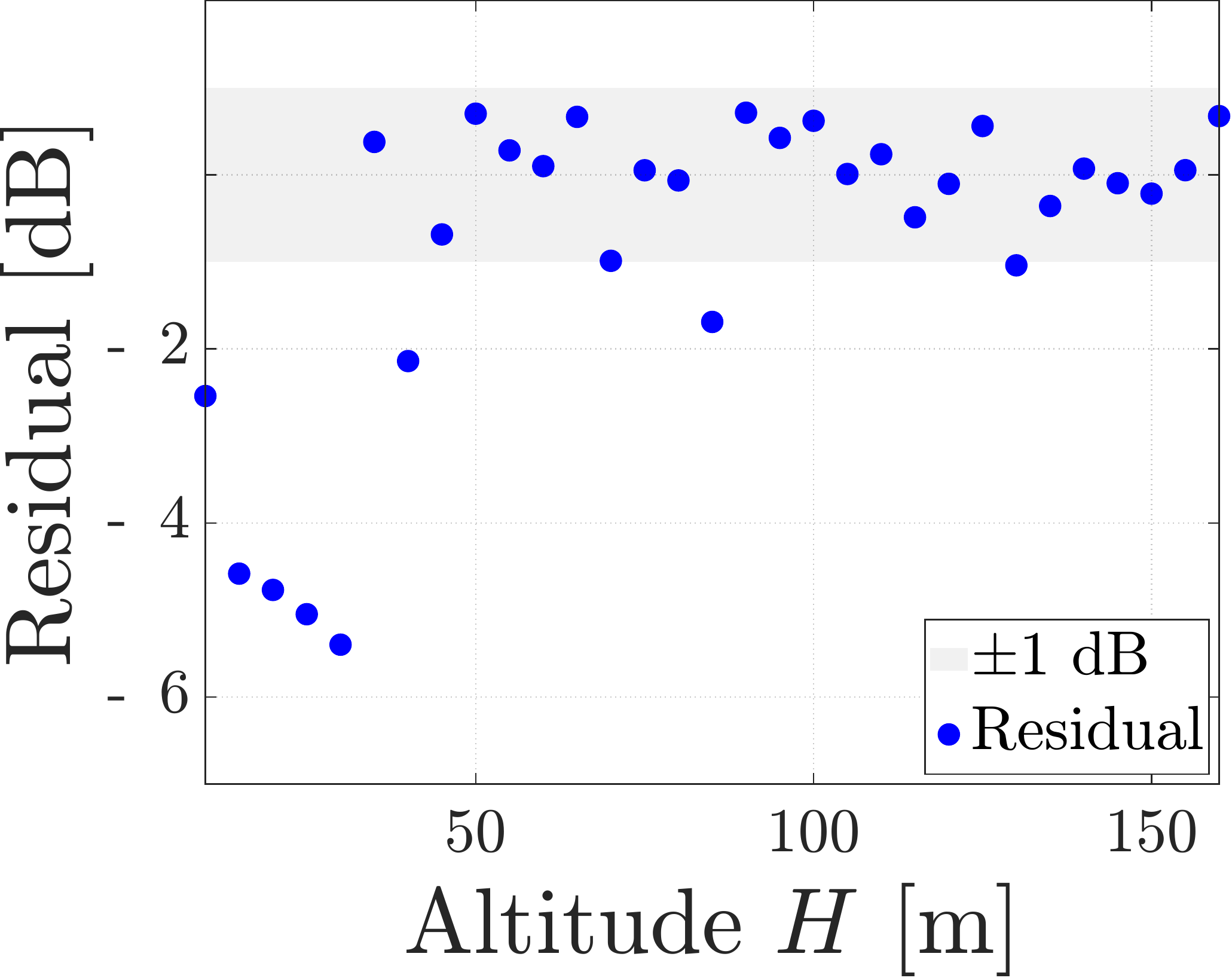}}
\hspace{0.01\linewidth}
\subfloat[5G n5 DL]{%
\includegraphics[width=\figW,height=\figH,keepaspectratio]{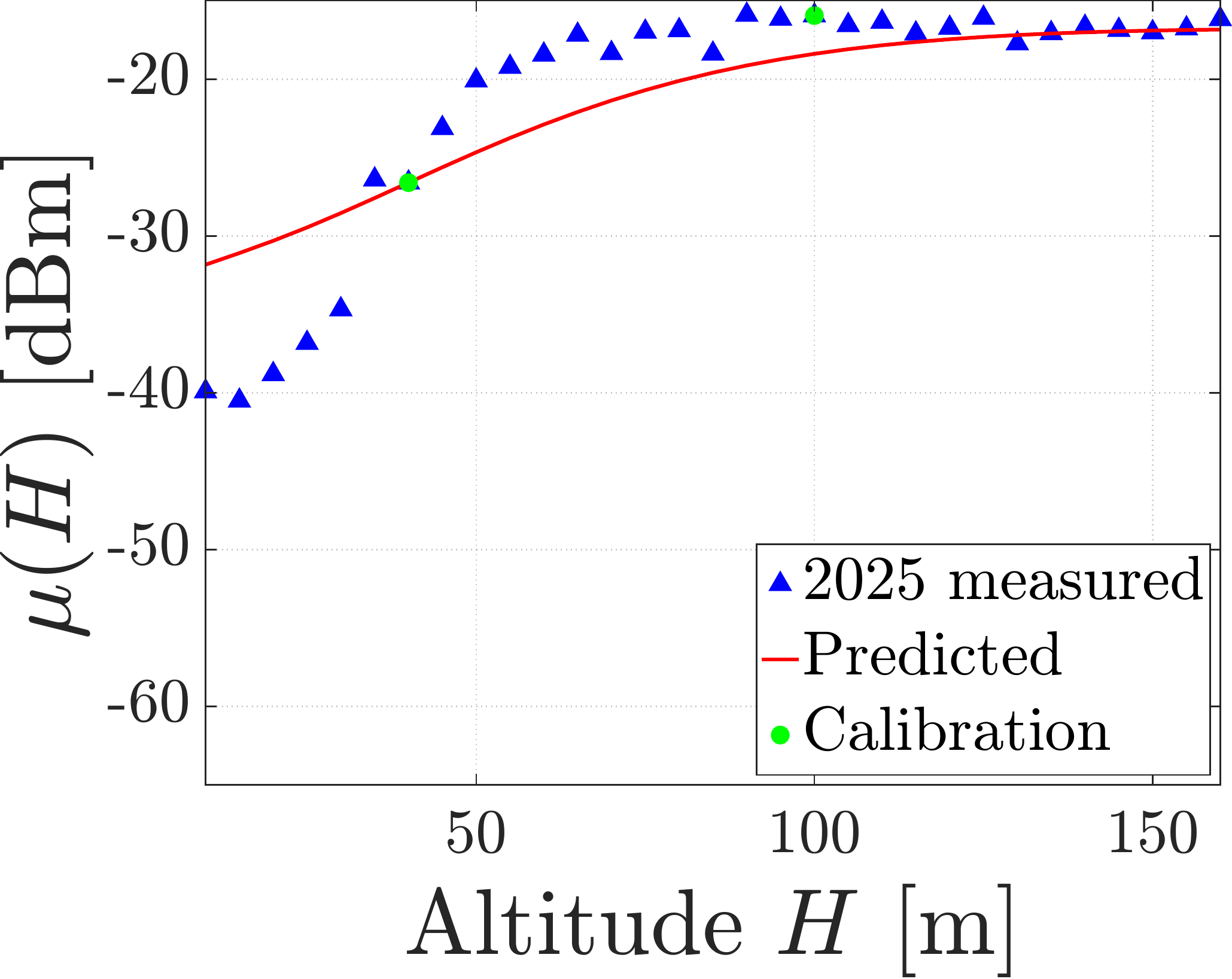}}\\

\subfloat[5G n5 UL]{%
\includegraphics[width=\figW,height=\figH,keepaspectratio]{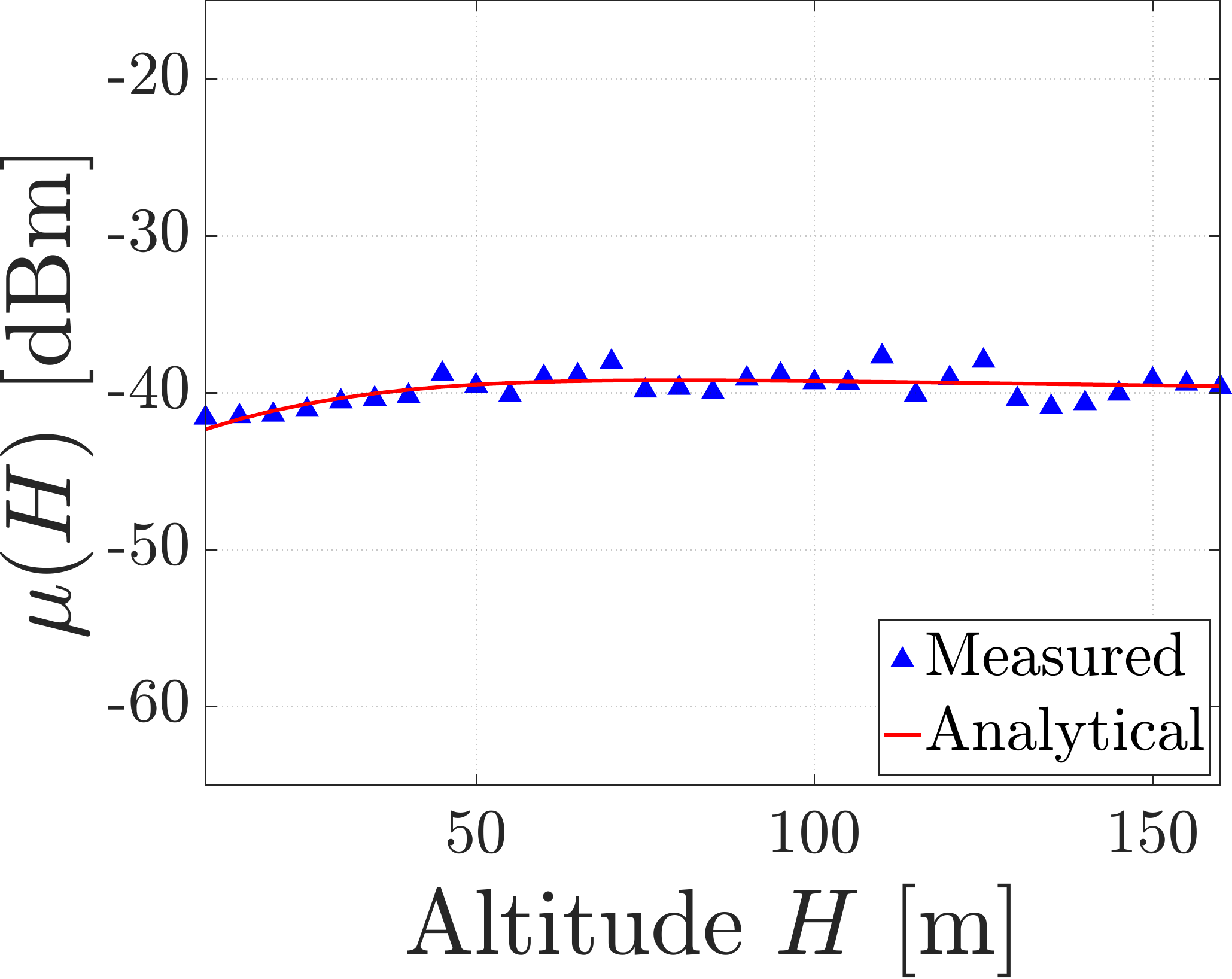}}
\hspace{0.01\linewidth}
\subfloat[5G n5 UL]{%
\includegraphics[width=\figW,height=\figH,keepaspectratio]{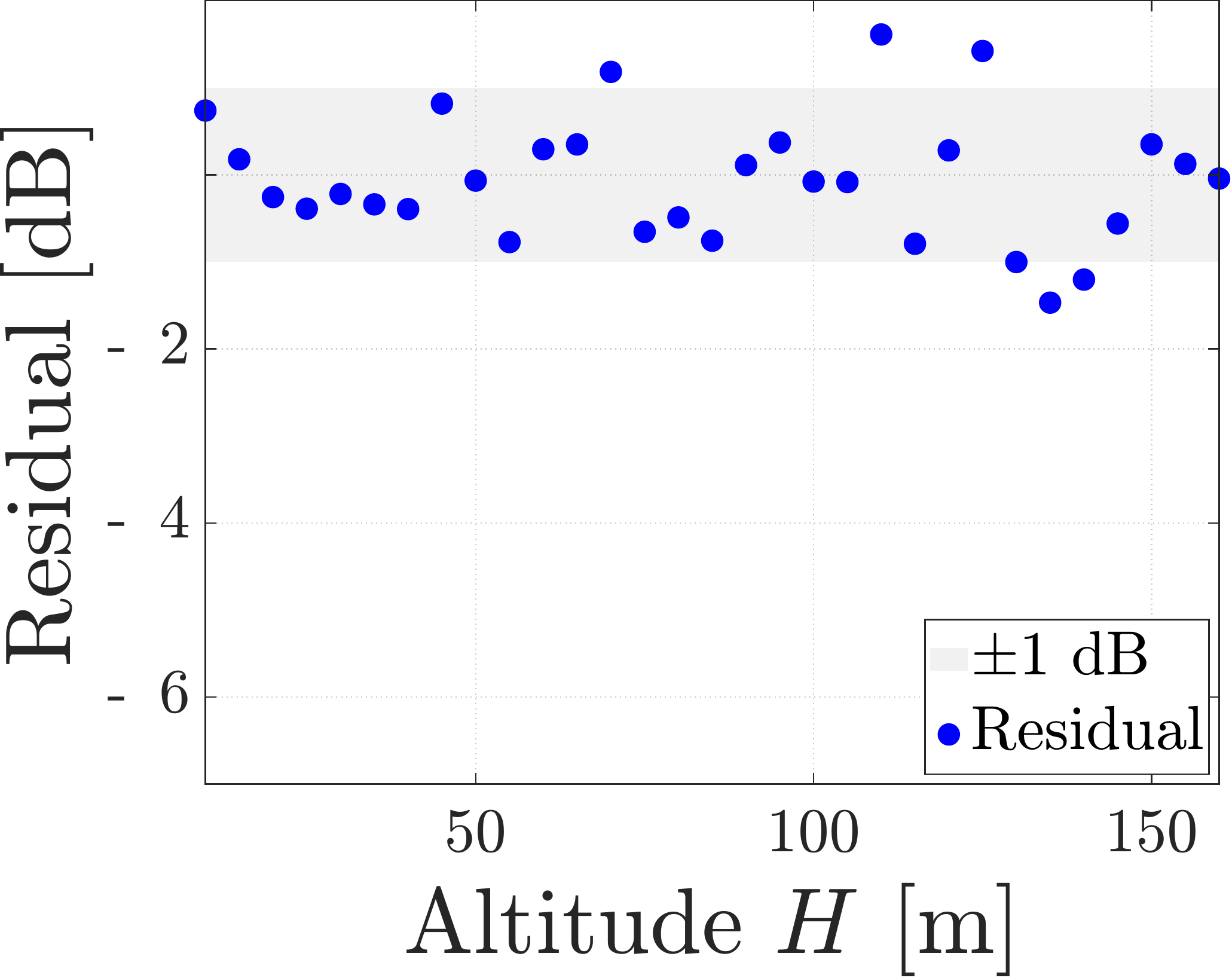}}
\hspace{0.01\linewidth}
\subfloat[5G n5 UL]{%
\includegraphics[width=\figW,height=\figH,keepaspectratio]{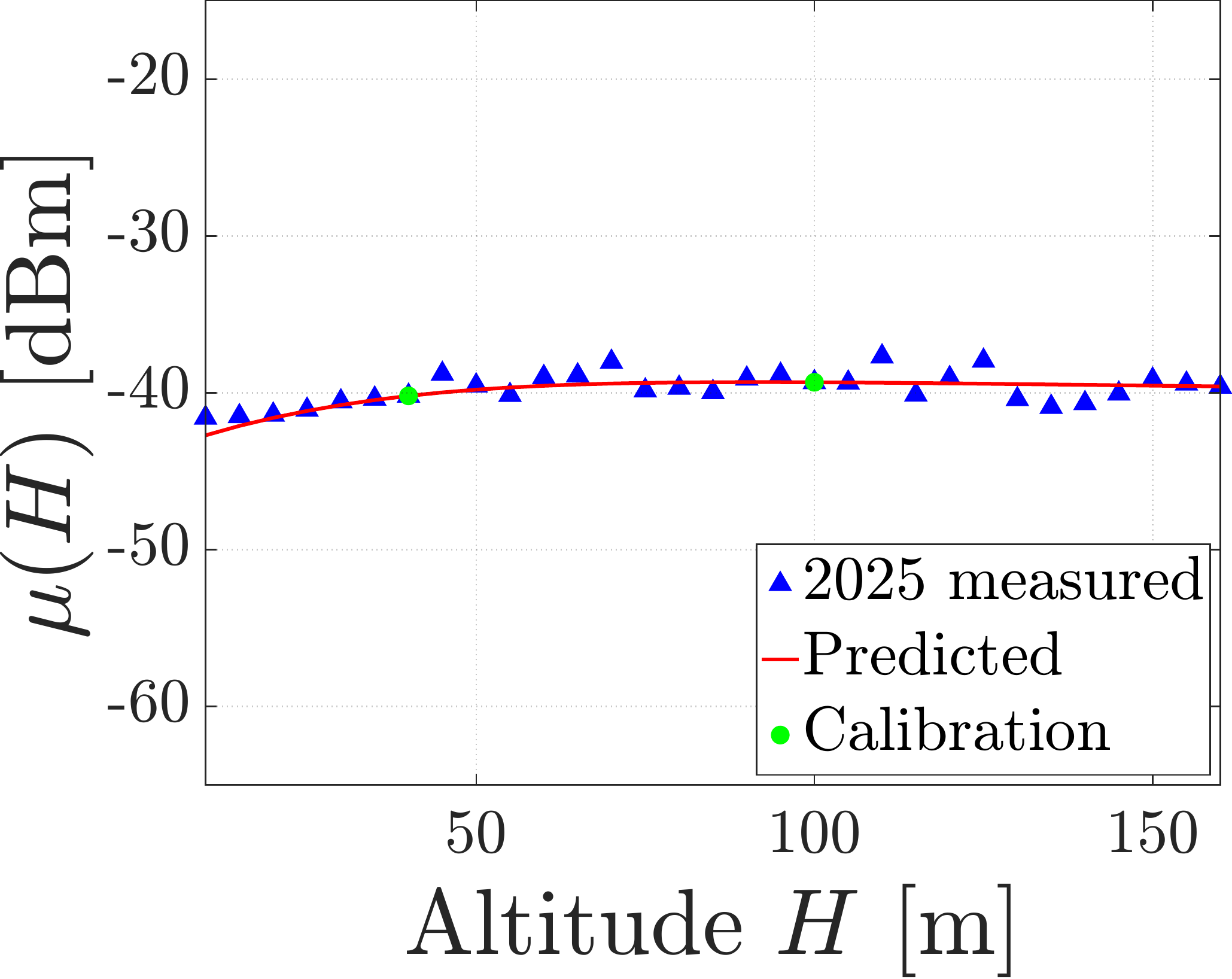}}\\

\subfloat[LTE Band~13 DL]{%
\includegraphics[width=\figW,height=\figH,keepaspectratio]{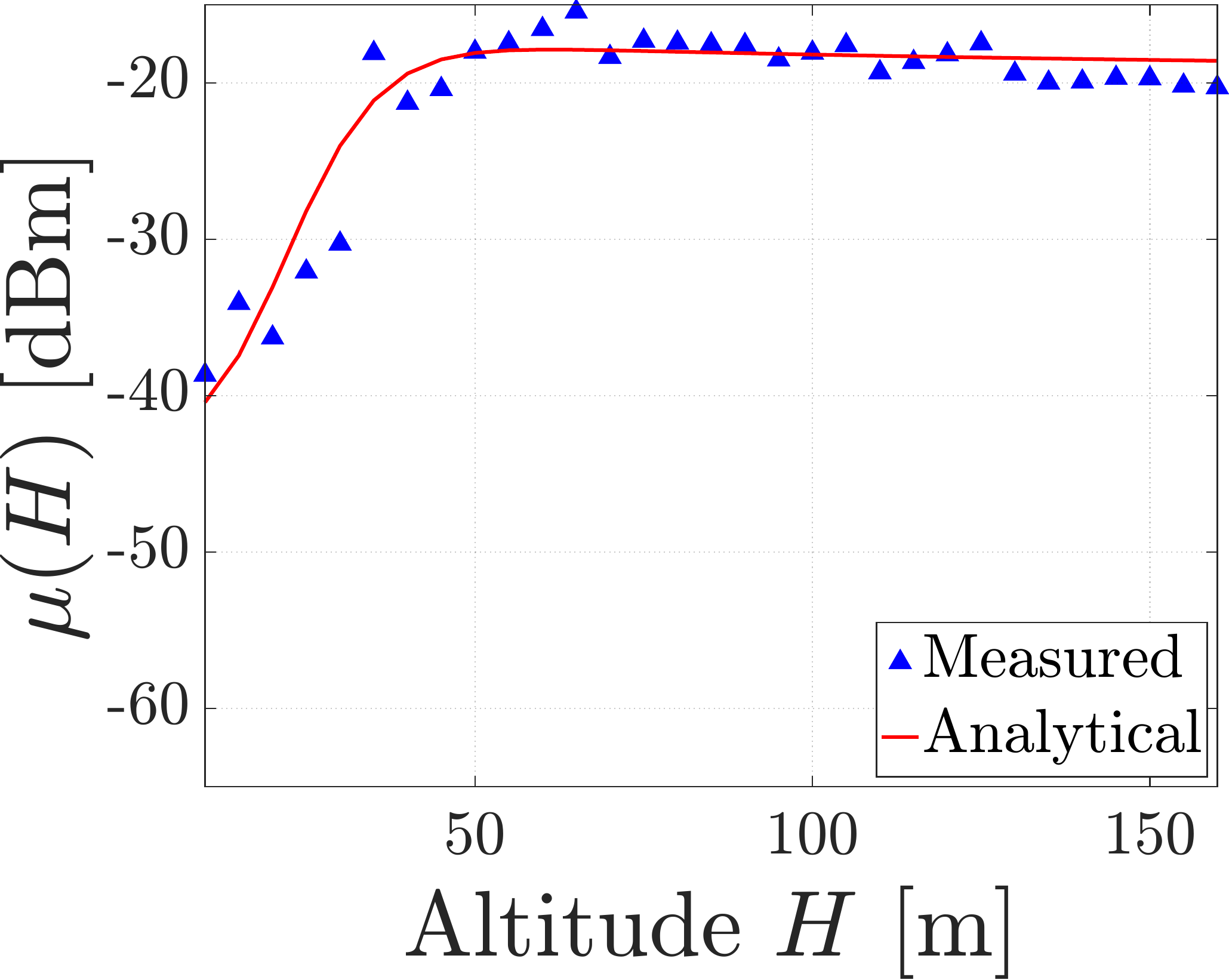}}
\hspace{0.01\linewidth}
\subfloat[LTE Band~13 DL]{%
\includegraphics[width=\figW,height=\figH,keepaspectratio]{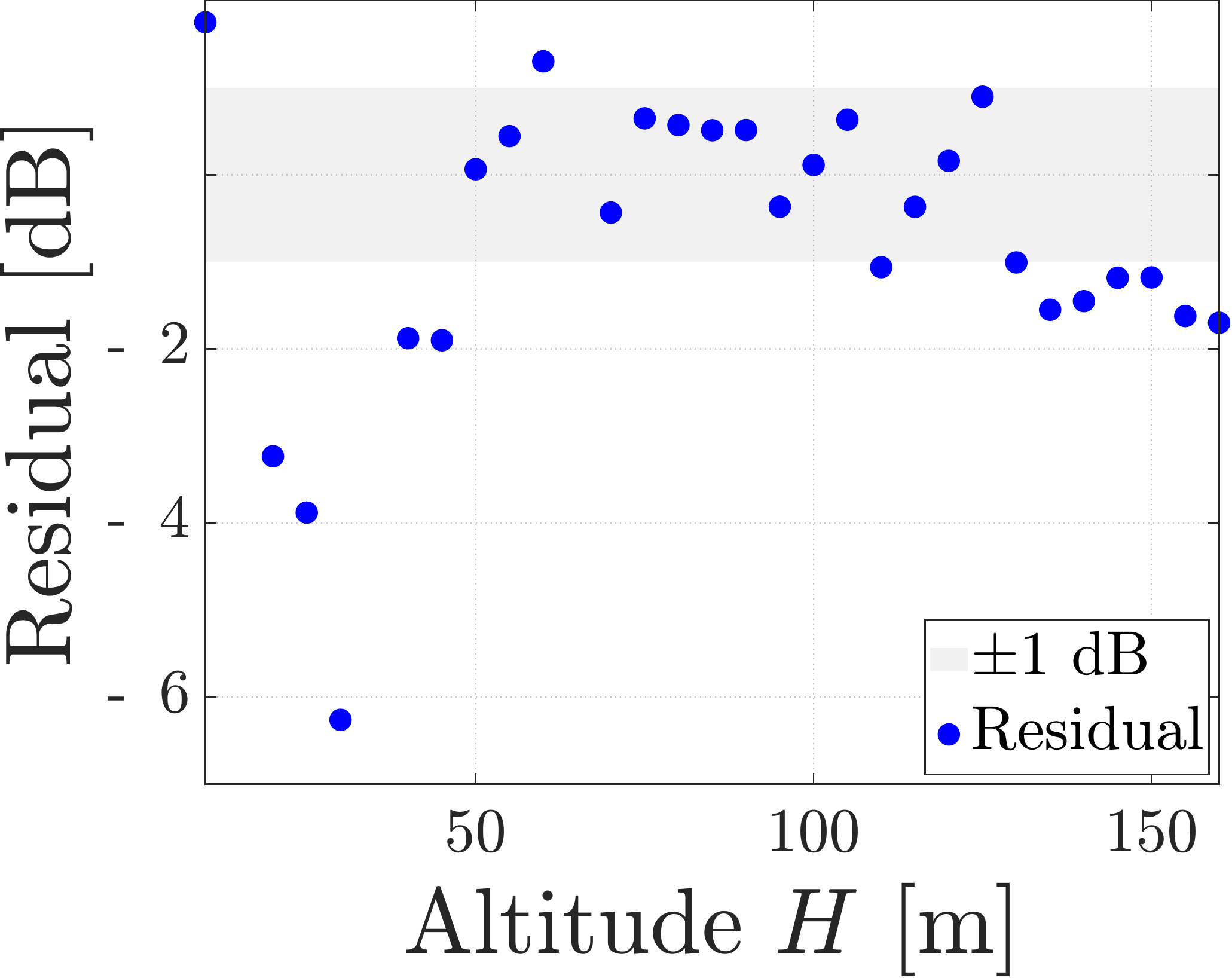}}
\hspace{0.01\linewidth}
\subfloat[LTE Band~13 DL]{%
\includegraphics[width=\figW,height=\figH,keepaspectratio]{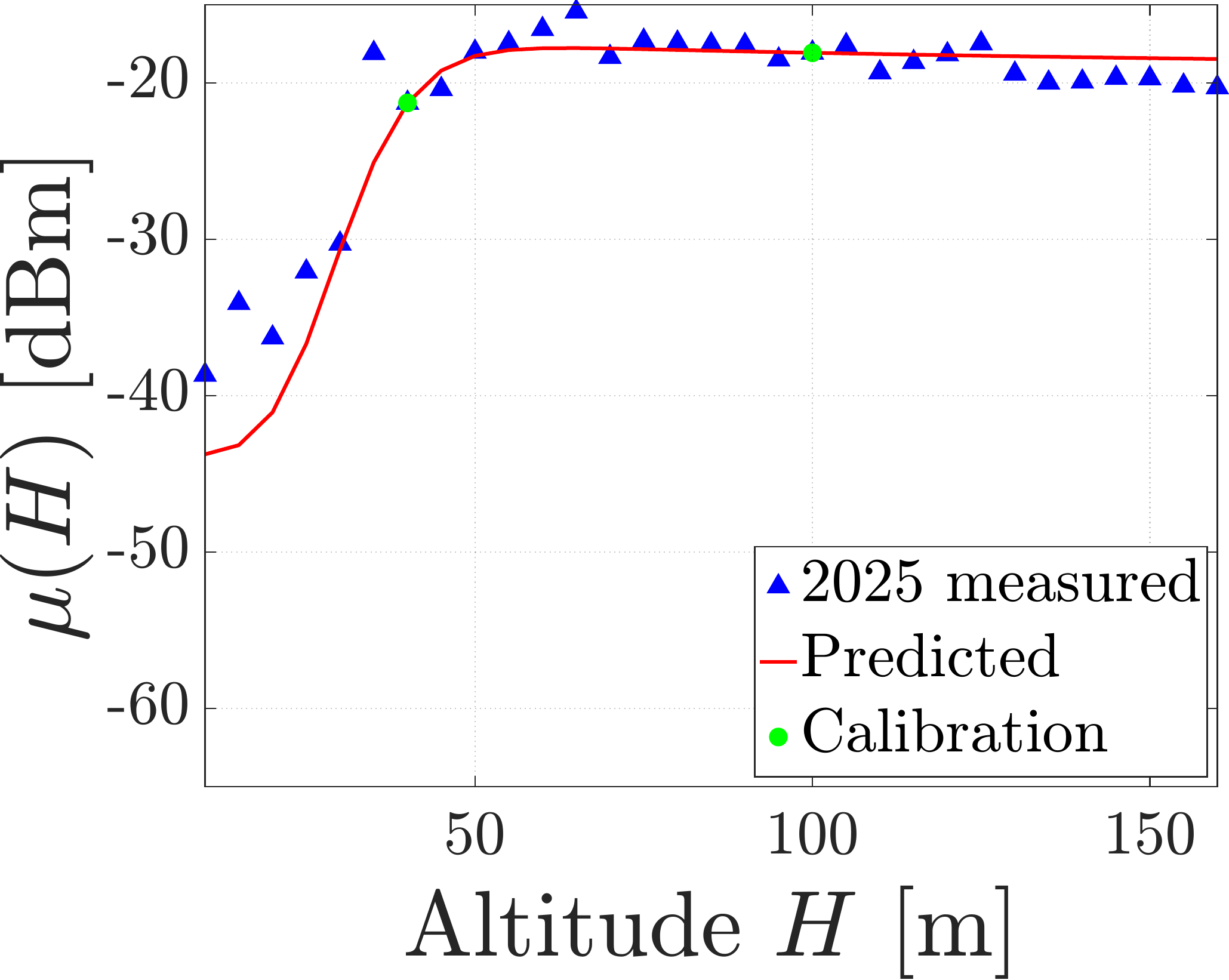}}\\

\subfloat[LTE Band~13 UL]{%
\includegraphics[width=\figW,height=\figH,keepaspectratio]{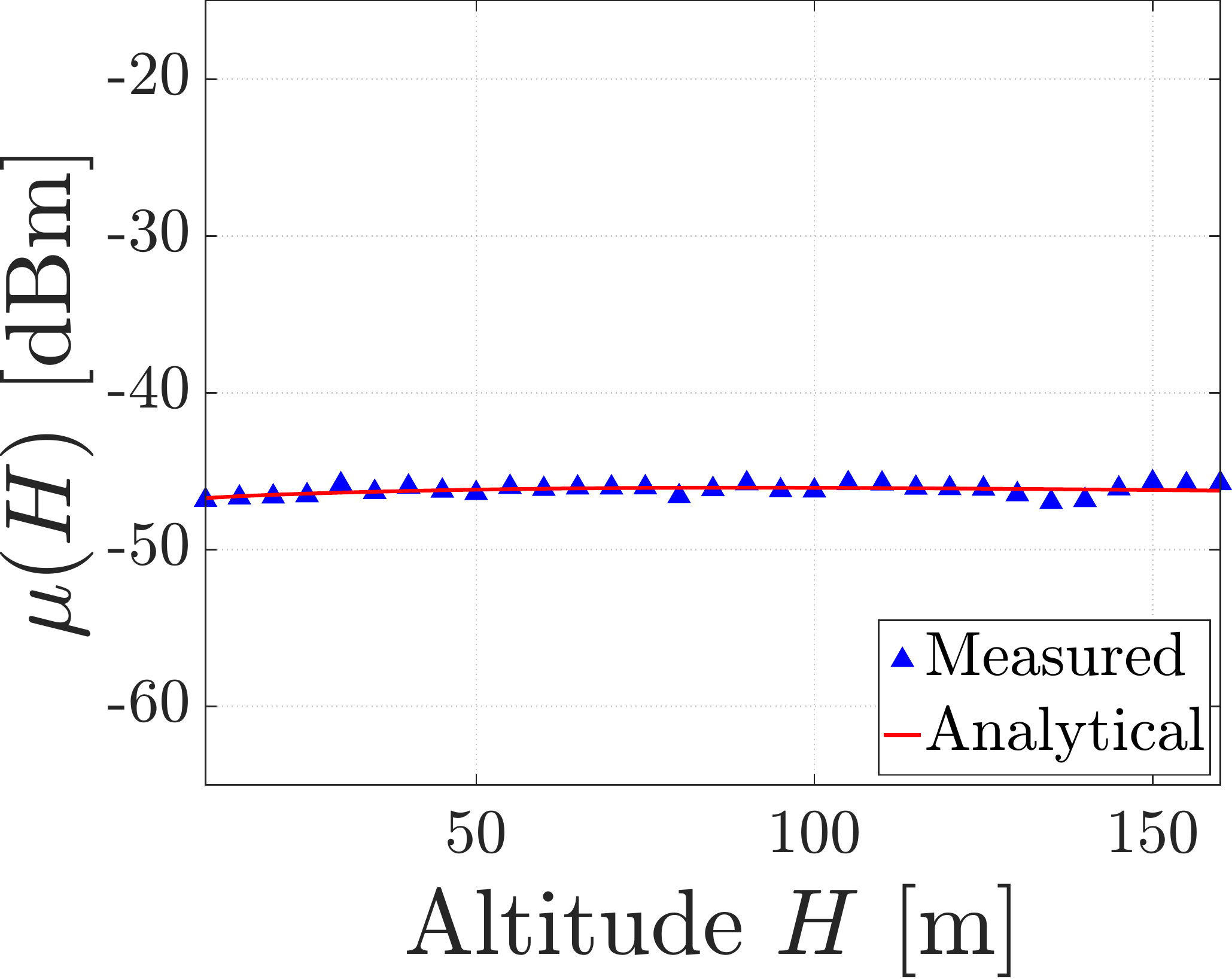}}
\hspace{0.01\linewidth}
\subfloat[LTE Band~13 UL]{%
\includegraphics[width=\figW,height=\figH,keepaspectratio]{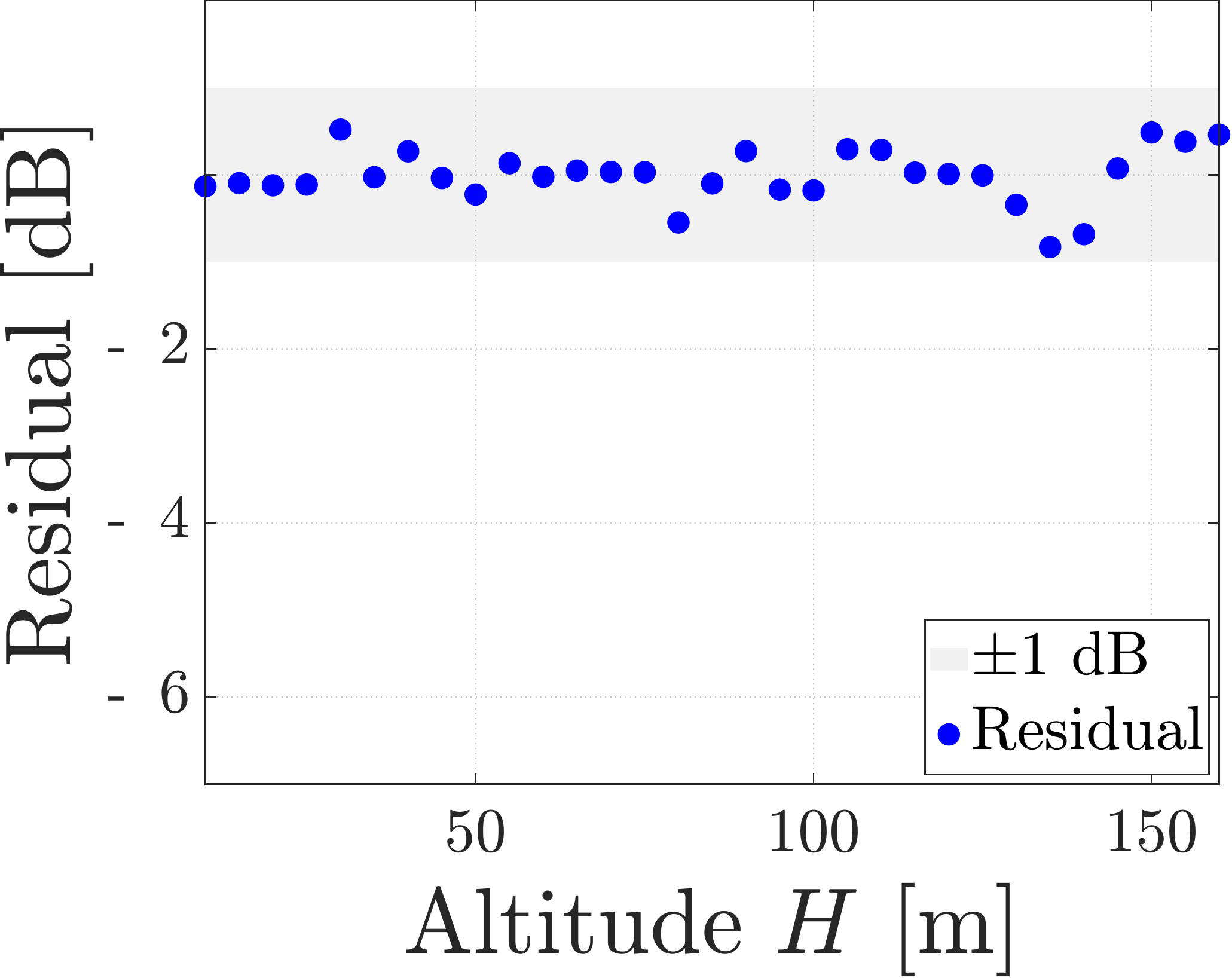}}
\hspace{0.01\linewidth}
\subfloat[LTE Band~13 UL]{%
\includegraphics[width=\figW,height=\figH,keepaspectratio]{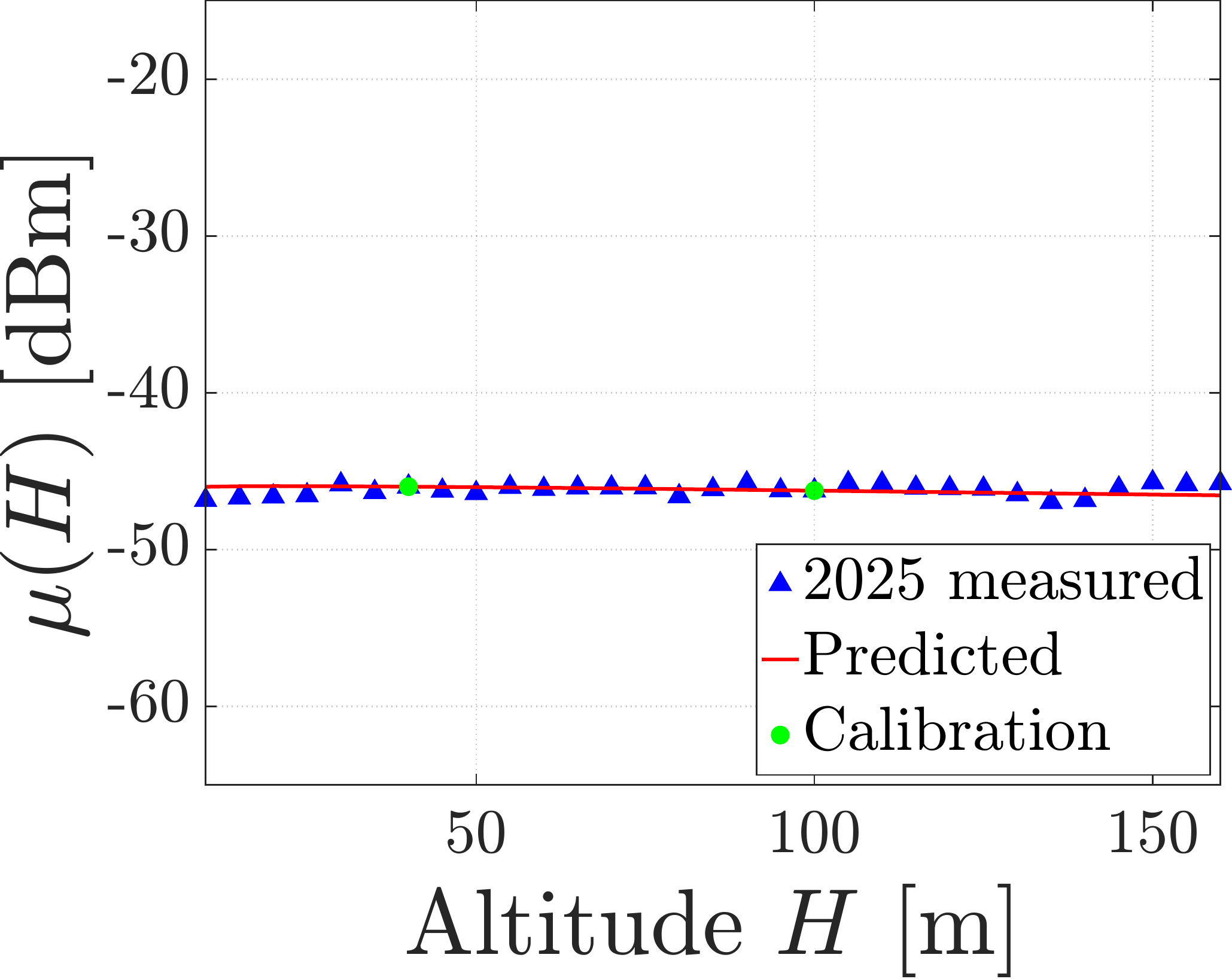}}\\

\subfloat[CBRS]{%
\includegraphics[width=\figW,height=\figH,keepaspectratio]{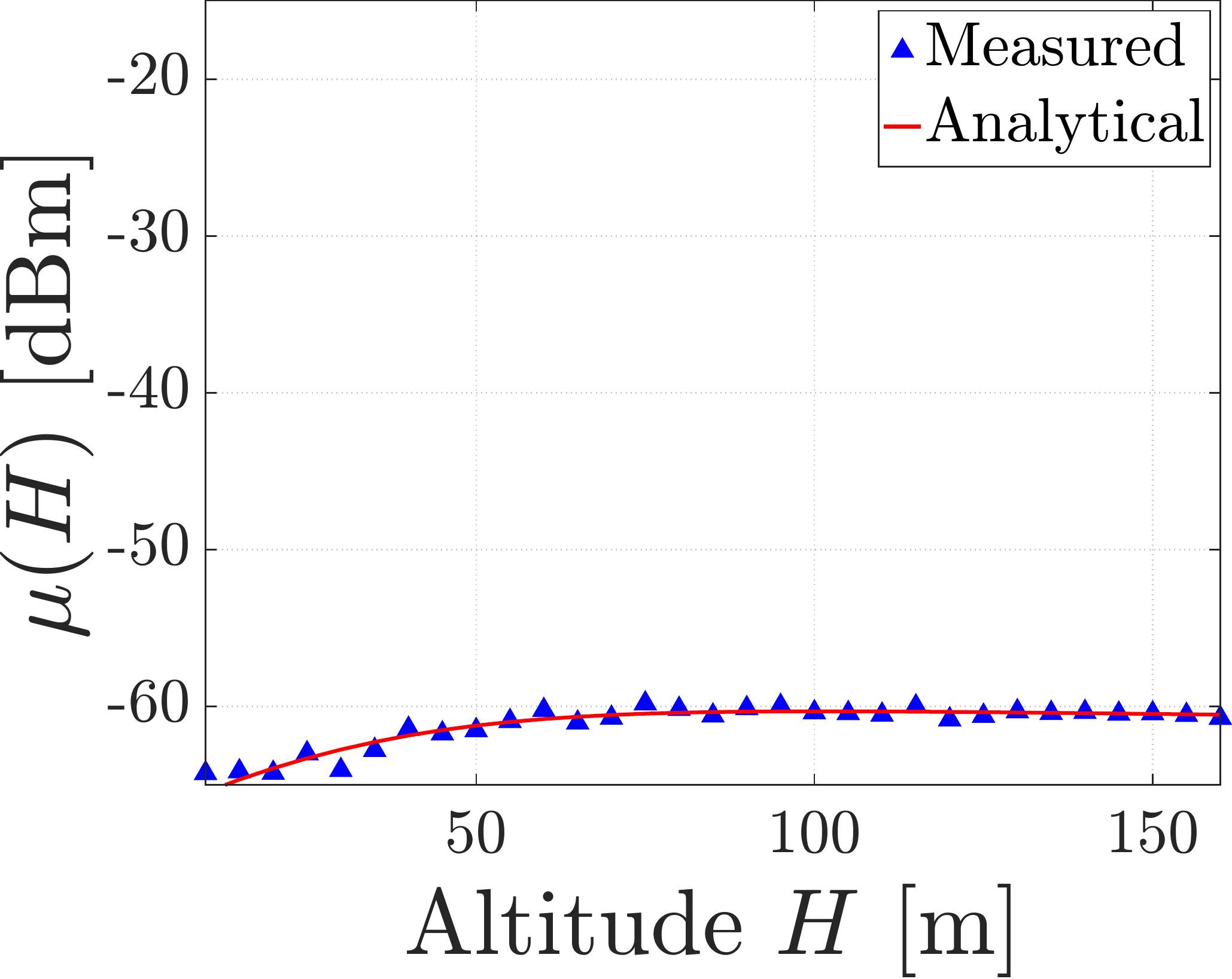}}
\hspace{0.01\linewidth}
\subfloat[CBRS]{%
\includegraphics[width=\figW,height=\figH,keepaspectratio]{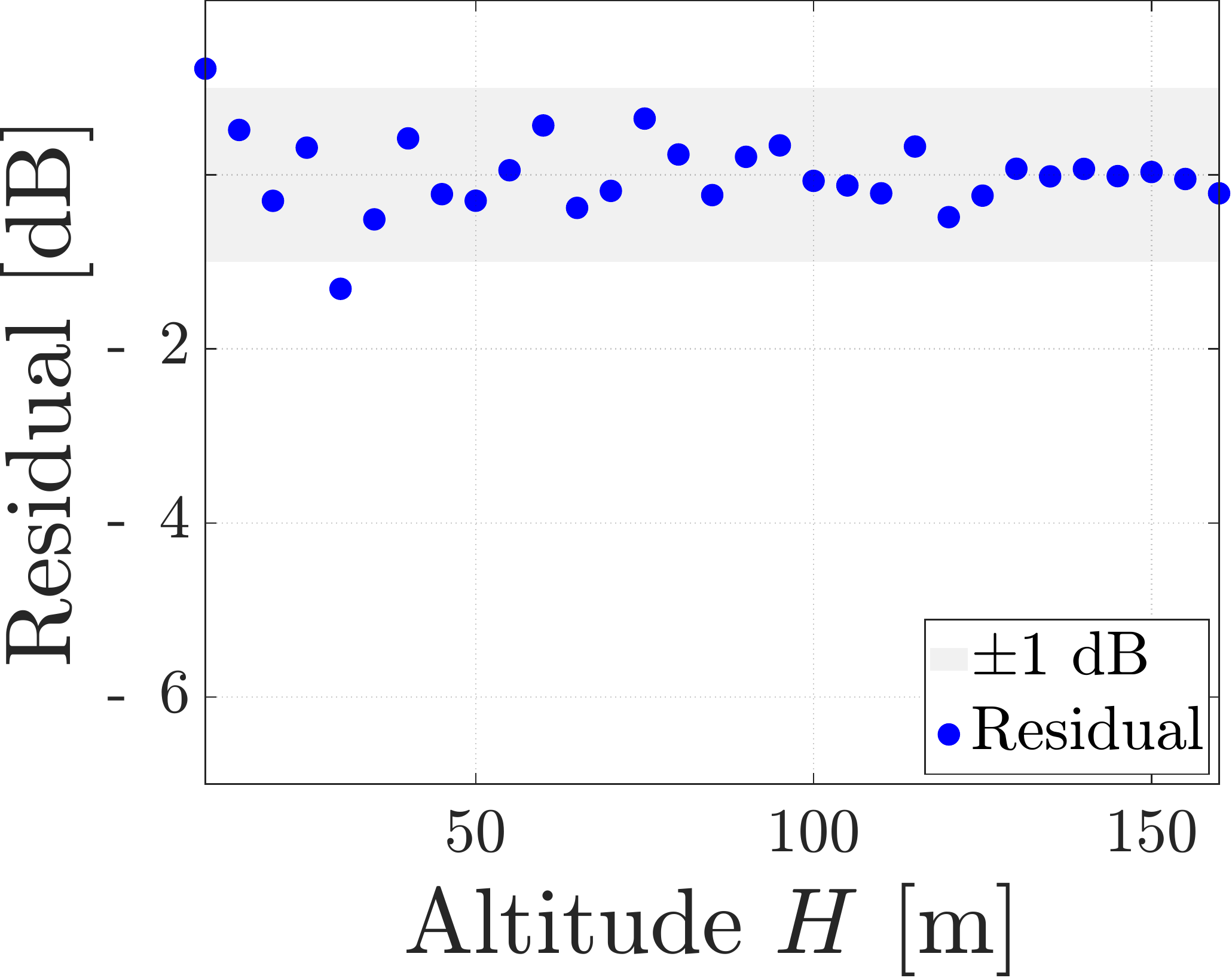}}
\hspace{0.01\linewidth}
\subfloat[CBRS]{%
\includegraphics[width=\figW,height=\figH,keepaspectratio]{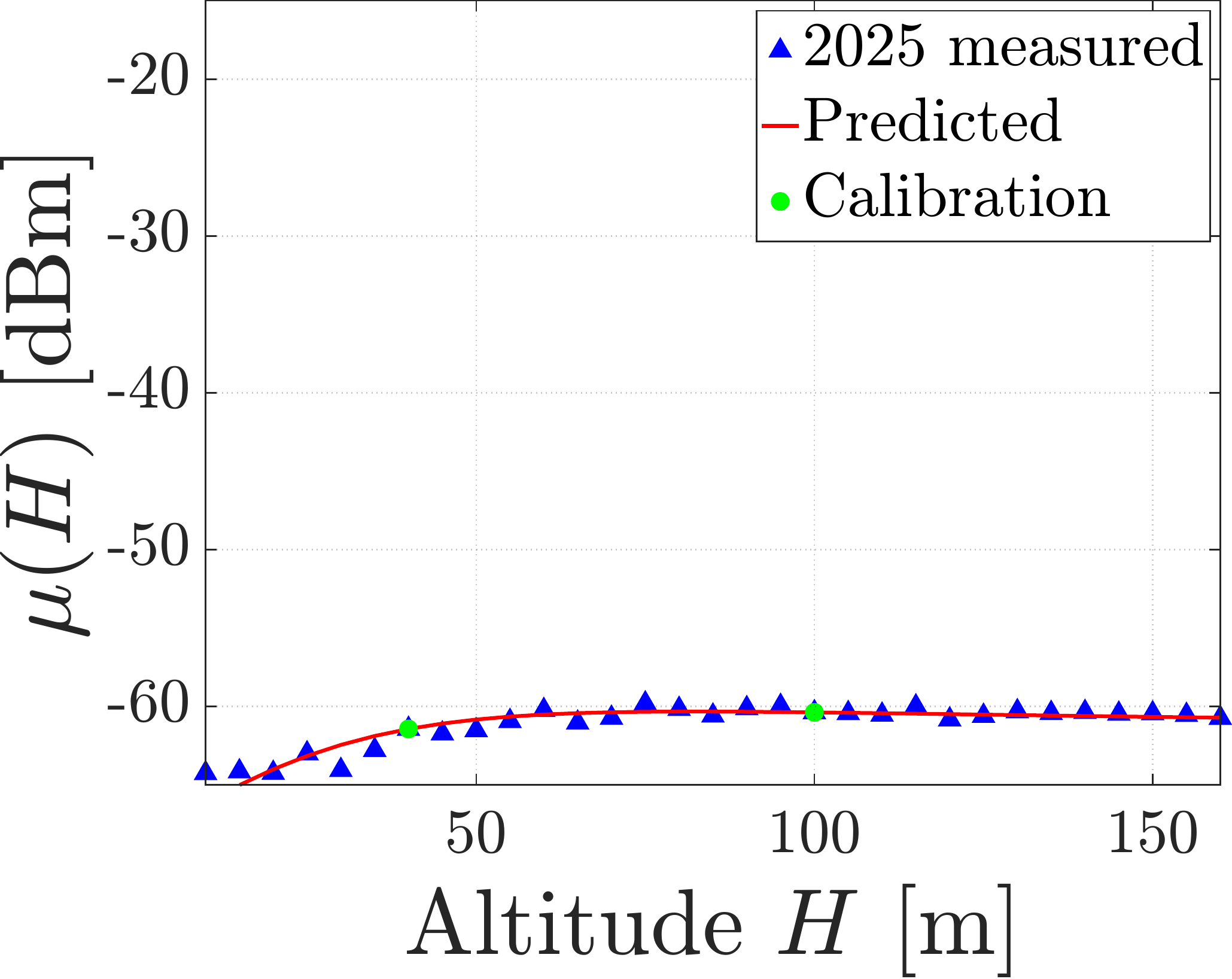}}

\caption{Altitude-dependent mean interference across multiple bands (2025). Left: measured mean interference (markers) and fitted mean-only model (solid lines) versus altitude. Middle: dB-domain residuals between measured and fitted means. Right: transferred mean interference models, with solid curves showing predictions with \(H_0\) fixed to 2024 and \((\beta, C_{\mathrm{eff}})\) calibrated from two 2025 altitude bins.}
\label{fig:residual_mean_interference_2025}
\end{figure*}

Table~\ref{tab:fit_summary} summarizes the fitted altitude-dependent model parameters and goodness-of-fit metrics across the bands and measurement campaigns. The goodness-of-fit metrics reported in this work capture complementary aspects of model performance. The linear-domain root mean squared error~(RMSE) quantifies absolute error in received power and is therefore dominated by high-power samples, making it sensitive to a small number of large deviations. Similarly, the linear coefficient of determination $R^2_{\mathrm{lin}}$ measures the fraction of variance explained in linear scale and can be strongly influenced by rare high-power events, potentially yielding high values even when the altitude-dependent trend is poorly represented. By contrast, the dB-domain coefficient of determination $R^2_{\mathrm{dB}}$ evaluates agreement on a logarithmic scale and is thus more sensitive to relative (multiplicative) variations and trend alignment over altitude, while being less affected by outliers. Given the mean-only, propagation-driven nature of the proposed model and its focus on capturing altitude-dependent structure rather than instantaneous power levels, dB-domain metrics are emphasized when assessing parameter identifiability and inter-year robustness.

Finally, the estimated frequency-normalized activity index $\tilde{C}$ provides physical insight into the operational hierarchies of the measured networks. As shown in Table~\ref{tab:fit_summary}, LTE Band~13 consistently exhibits the highest values of $\tilde{C}$ ($\approx 10^{-5}$), aligning with expectations for a high-power, licensed band with extensive macrocell deployment. In comparison, the 5G n5 band yields intermediate values ($\approx 10^{-6}$), reflecting varying deployment densities and the potential sharing of spectrum with legacy LTE services. Notably, the CBRS band exhibits parameter values several orders of magnitude smaller~($\approx 10^{-11}$), consistent with its power-limited, shared-access nature and predominant use in small-cell architectures. Furthermore, uplink~(UL) entries across all bands consistently show lower $\tilde{C}$ values than their downlink~(DL) counterparts, attributable to lower user transmit power and sporadic transmission patterns. These distinct stratifications confirm that $\tilde{C}$ effectively separates intrinsic network activity from frequency-dependent propagation effects, making it a robust metric for cross-band characterization.

For the DL bands (LTE Band~13 DL and 5G~n5 DL), the fitted parameters are stable and physically interpretable across years. The transition altitude $H_0$ consistently lies within the measured altitude range (approximately $15$-$30$~m), and the transition slope $\beta$ takes moderate values, indicating a resolvable but gradual change in altitude dependence. These bands also exhibit high goodness of fit in the dB domain, demonstrating that the mean-only model accurately captures the dominant altitude trend. The comparatively lower $R^2_{\mathrm{lin}}$ values reflect sensitivity to a small number of high-power samples in linear scale rather than systematic modeling error. Overall, these results indicate that, for DL bands, altitude-dependent geometric effects constitute a dominant and persistent component of the aggregate interference.

In contrast, UL bands (LTE Band~13 UL and 5G~n5 UL) consistently yield very small transition slopes and large negative values of $H_0$. These negative $H_0$ values are not physically meaningful as altitudes; instead, they indicate that no distinct LoS transition is identifiable within the observed altitude range. In this regime, the logistic transition effectively degenerates to a nearly constant function, and $\beta$ becomes weakly identifiable. This interpretation is supported by the consistently low $R^2_{\mathrm{dB}}$ values for UL bands, despite moderate $R^2_{\mathrm{lin}}$ in some cases, indicating that apparent linear-scale agreement is driven by overall power level rather than by accurate modeling of altitude-dependent structure. Practically, this implies that mean-only altitude-dependent models provide limited explanatory power for UL interference, where activity variability dominates over geometric propagation effects.

The CBRS band exhibits a distinct intermediate behavior. While $H_0$ varies across years and  takes negative values, the fitted transition slopes remain small and stable, and the goodness-of-fit metrics are consistently high in both linear and dB domains. This suggests that CBRS interference exhibits a smooth, weakly altitude-dependent mean that is well captured by the model without a sharp transition. 

Fig.~\ref{fig:residual_mean_interference_2025}, left column, compares the 2025 altitude-binned mean interference (i.e., $10\log_{10}\!\langle I(h)\rangle$) shown in markers against the fitted mean-only model (solid lines). The corresponding residuals in dB are reported in the middle column. A \(\pm 1\)~dB band is shown as a visual reference. Three distinct behaviors are evident. First, the two DL bands (5G n5 DL and LTE Band~13 DL) exhibit a pronounced low-altitude rise followed by a saturation (near-constant mean) at moderate-to-high altitudes. In both cases, the model captures the overall S-shaped transition and the high-altitude plateau; however, the residuals reveal a systematic mismatch at the lowest altitude bins. Specifically, the DL residuals reach approximately \(-5\) to \(-7\)~dB at the smallest altitudes, then collapse toward near-zero residuals beyond roughly the transition region. This pattern indicates that the fitted curve is consistently above the measured mean at very low altitudes, while it becomes accurate over the remainder of the altitude range. 

Second, the UL bands (5G n5 UL and LTE Band~13 UL) are nearly altitude-invariant in the mean, with the fitted curves closely tracking the measured means across the full altitude range. This outcome reflects the low elevation and clutter-embedded nature of user equipment, which lacks the geometric separation required to produce a resolvable line-of-sight transition. As a result, the mean interference profile remains effectively flat with altitude. This altitude invariance is further reinforced by closed-loop power control and bursty transmission activity, both of which introduce variability that obscures any underlying geometric trend. Correspondingly, their residuals remain tightly clustered around 0~dB and largely within the $\pm1$~dB band, without the pronounced low-altitude deviation observed in the DL.

Third, CBRS exhibits a smooth mean profile with a gradual increase followed by saturation, reflecting both its deployment characteristics and duplexing structure. Specifically, low-power shared-access nodes operating at or near rooftop height generate a diffuse interference field that lacks the sharp geometric transition observed in macrocell DL bands. This behavior is further shaped by Time Division Duplexing~(TDD), which aggregates both UL and DL transmissions into the mean measurement. Since high-power DL slots dominate the aggregate energy, the measured mean retains a mild altitude-dependent trend characteristic of base station interference, while the temporal mixing of user traffic acts to smooth residual structure across altitudes.


Fig.~\ref{fig:residual_mean_interference_2025}, right column, illustrates the inter-year transfer of altitude-dependent mean interference models from 2024 to 2025. Fixing $H_0$ to its 2024 value and calibrating $(\beta, C_{\mathrm{eff}})$ using only two altitude bins in 2025 imposes a structural constraint on the model. For DL bands, where the fitted $H_0$ remains consistently positive, the transferred model successfully reproduces the altitude trend beyond the calibration region. This suggests that despite parameter drift, the functional form of the altitude profile remains robust enough for limited transfer.

Conversely, UL bands exhibit large inter-year variability and frequent sign changes in $H_0$, indicating the lack of a stable transition structure. Consequently, fixing $H_0$ from a prior year yields transferred curves that are weakly constrained, resulting in poor goodness-of-fit outside the calibration bins (e.g., negative $R^2_{\mathrm{dB}}$ for LTE B13 UL). In these cases, the ``transfer failure'' is attributable to the absence of a persistent altitude-dependent structure rather than the calibration method itself. The CBRS band occupies an intermediate regime: while $H_0$ varies, the smooth monotonic interference profile allows the two-point calibration to recover a reasonable approximation of the 2025 data. For bands with limited dynamic range ($\approx$ 4--5~dB), the exact transition location is weakly identifiable; thus, large variations in $H_0$ have negligible impact on prediction accuracy.

Table~\ref{tab:transfer_40_100} summarizes the transfer performance. The \textit{Transfer Score} is defined as the ratio of the transfer RMSE to the RMSE of a direct fit on 2025 data; values near 1.0 indicate that the transfer method performs comparably to a full dataset fit. DL and CBRS bands maintain moderate absolute error (RMSE$_{\mathrm{dB}} < 4$~dB) and positive $R^2_{\mathrm{dB}}$. The transfer scores range from 1.34 to 1.99, indicating that while transfer introduces some error degradation compared to fitting the full 2025 dataset, the model remains predictive. UL bands show low or negative $R^2_{\mathrm{dB}}$ despite low absolute RMSE, reflecting the instability of $R^2$ when the underlying response variance is small.

\begin{table}[!t]
\centering
\caption{Inter-year transfer performance from 2024 to 2025 using two-point altitude calibration at 40~m and 100~m.}
\label{tab:transfer_40_100}
\scalebox{0.75}{
\begin{tabular}{l c c c c c c}
\toprule
\textbf{Band} &
$\boldsymbol{\beta}$ &
$\boldsymbol{\tilde{C}}$ &
\textbf{RMSE}$_{\mathrm{dB}}$ &
$\mathbf{R^2_{\mathrm{dB}}}$ &
$\boldsymbol{\Delta}$\textbf{RMSE}$_{\mathrm{dB}}$ &
\textbf{Score} \\
\midrule
LTE B13 DL & 0.19 & 1.28 & 2.77 & 0.79 & +0.79 & 1.40 \\
LTE B13 UL & 0.03 & $2.17{\times}10^{-3}$ & 0.43 & $-0.57$ & +0.12 & 1.40 \\
5G n5 DL   & 0.04 & 2.69 & 3.89 & 0.76 & +1.93 & 1.99 \\
5G n5 UL   & 0.05 & $1.19{\times}10^{-2}$ & 0.75 & 0.42 & +0.04 & 1.05 \\
CBRS       & 0.06 & $1.72{\times}10^{-3}$ & 0.59 & 0.81 & +0.15 & 1.34 \\
\bottomrule
\end{tabular}
}
\end{table}

The transfer score is defined as the ratio between the RMSE of the transferred model and the RMSE obtained from a direct 2025 fit, such that values near one denote comparable performance and values above one indicate degradation due to transfer. DL and shared-access bands (LTE B13 DL, 5G n5 DL, CBRS) preserve strong altitude-dependent structure under inter-year transfer, exhibiting moderate absolute error (RMSE$_{\mathrm{dB}} < 4$~dB), positive $R^2_{\mathrm{dB}}$, and transfer scores close to unity, indicating limited degradation relative to direct fitting using 2025 data. In contrast, UL bands exhibit weaker or inconsistent altitude dependence, which can lead to low or negative $R^2_{\mathrm{dB}}$ despite small RMSE$_{\mathrm{dB}}$, reflecting limited identifiability of altitude-transition structure rather than calibration failure; for such bands, RMSE is a more informative metric than $R^2$, which becomes unstable when the response variance is small.

\section{Conclusion}
This paper investigated the inter-year transferability of altitude-dependent mean interference models using aerial spectrum measurements. Using a mean-only stochastic-geometry framework, we demonstrated that altitude-dependent structure is captured accurately and consistently for DL-dominated and shared-access bands (including CBRS), where fitted parameters remain stable and physically interpretable across years. Conversely, UL bands consistently exhibit weak altitude dependence with no identifiable transition, limiting the applicability of geometric mean-field models.
Inter-year reuse was evaluated through a minimal two-point transfer calibration from 2024 to 2025, in which the transition altitude $H_0$ was fixed to the reference year value, and only the transition slope $\beta$ and effective activity factor $C_{\mathrm{eff}}$ were estimated. Results were strongly band-dependent. For DL and shared-access bands, the transferred models successfully preserved the dominant altitude trend with only moderate degradation, confirming that geometry-driven propagation effects are persistent and reusable. However, transfer performance for UL bands was dominated by systematic bias, reflecting the fundamental absence of a stable altitude structure rather than deficiencies in the calibration procedure.
Overall, these findings indicate that inter-year reuse of altitude-dependent mean models is viable only when propagation effects dominate aggregate interference behavior. Practically, this suggests that a sparse set of calibration measurements may be sufficient for DL and shared-access bands in stable environments. This capability can significantly reduce the duration of measurement campaigns and enable the use of narrowband spectrum sensors, which are otherwise constrained by sweep latency~\cite{cullen2023predicting}. On the other hand, UL interference modeling likely requires denser recalibration or alternative formulations that explicitly account for traffic variability and power-control dynamics.

\bibliographystyle{IEEEtran}
\bibliography{references}

@article{maeng2025altitude,
  title={Altitude-dependent sub-6 {GHz} wireless spectrum: {Survey}, measurements, trends, and modeling},
  author={Maeng, Sung Joon and Raouf, Amir Hossein Fahim and Ozdemir, Ozgur and Zajkowski, Thomas and Mushi, Magreth and Sichitiu, Mihail L and Dutta, Rudra and Guvenc, Ismail},
  journal={Authorea Preprints},
  year={2025},
  publisher={Authorea}
}

@article{Mozaffari2019Tutorial,
  title={A tutorial on {UAVs} for wireless networks: {Applications}, challenges, and open problems},
  author={Mozaffari, Mohammad and Saad, Walid and Bennis, Mehdi and Nam, Young-Han and Debbah, M{\'e}rouane},
  journal={IEEE Commun. Surv. Tutor.},
  volume={21},
  number={3},
  pages={2334--2360},
  year={2019},
  publisher={IEEE}
}

@ARTICLE{Fotouhi2019Survey,
  author={Fotouhi, Azade and Qiang, Haoran and Ding, Ming and Hassan, Mahbub and Giordano, Lorenzo Galati and Garcia-Rodriguez, Adrian and Yuan, Jinhong},
  journal={IEEE Commun. Surv. Tutor.}, 
  title={Survey on {UAV} Cellular Communications: {Practical} Aspects, Standardization Advancements, Regulation, and Security Challenges}, 
  year={2019},
  volume={21},
  number={4},
  pages={3417-3442},
  doi={10.1109/COMST.2019.2906228}
}

@article{AlHourani2014Optimal,
  title={Optimal {LAP} altitude for maximum coverage},
  author={Al-Hourani, Akram and Kandeepan, Sithamparanathan and Lardner, Simon},
  journal={IEEE Wirel. Commun.},
  volume={3},
  number={6},
  pages={569--572},
  year={2014},
  publisher={IEEE}
}

@article{Azari2019Cellular,
  title={Cellular connectivity for {UAVs}: {Network} modeling, performance analysis, and design guidelines},
  author={Azari, M Mahdi and Rosas, Fernando and Pollin, Sofie},
  journal={IEEE Trans. Wirel. Commun.},
  volume={18},
  number={7},
  pages={3366--3381},
  year={2019},
  publisher={IEEE}
}

@book{Haenggi2012SG,
  title={Stochastic geometry for wireless networks},
  author={Haenggi, Martin},
  year={2013},
  publisher={Cambridge University Press}
}

@article{matolak2017air,
  title={Air--ground channel characterization for unmanned aircraft systems—{Part} {III}: {The} suburban and near-urban environments},
  author={Matolak, David W and Sun, Ruoyu},
  journal={IEEE Trans. Veh. Technol.},
  volume={66},
  number={8},
  pages={6607--6618},
  year={2017},
  publisher={IEEE}
}

@article{Chu2019Interference,
  title={Interference modeling and analysis in 3-dimensional directional {UAV} networks based on stochastic geometry},
  author={Chu, Eunmi and Kim, Jong Min and Jung, Bang Chul},
  journal={ICT Express},
  volume={5},
  number={4},
  pages={235--239},
  year={2019},
  publisher={Elsevier}
}

@article{Selim2025UAV,
  title={Stochastic geometry analysis of {UAV-assisted} networks with probabilistic {UA} activation},
  author={Selim, Mahmoud M},
  journal={Sci. Rep.},
  volume={15},
  number={1},
  pages={37356},
  year={2025},
  publisher={Nature Publishing Group UK London}
}

@article{cullen2023predicting,
  title={Predicting dynamic spectrum allocation: {A} review covering simulation, modelling, and prediction},
  author={Cullen, Andrew C and Rubinstein, Benjamin IP and Kandeepan, Sithamparanathan and Flower, Barry and Leong, Philip HW},
  journal={Artif. Intell. Rev.},
  volume={56},
  number={10},
  pages={10921--10959},
  year={2023},
  publisher={Springer}
}
\end{document}